\documentstyle[11pt]{article}
\begin{document}
\sf
{

\vspace*{3.6cm}

\begin{center}
{\sc \bf
THEORETICAL NEUTRON-CAPTURE CROSS SECTIONS FOR R-PROCESS NUCLEOSYNTHESIS
IN THE $^{48}$CA REGION}

\bigskip

\bigskip

\underline{T. Rauscher}\footnote{Alexander von Humboldt fellow}, 
W. B\"ohmer, K.-L. Kratz\\
Institut f\"ur Kernchemie, Fritz-Strassmann-Weg 2, D-55099 Mainz, 
Germany\\
W. Balogh, H. Oberhummer\\
Institut f\"ur Kernphysik, Wiedner Hauptstr.\ 8--10, A-1040 Vienna, 
Austria

\vspace*{3cm}

Abstract:
\end{center}


We calculate neutron capture cross sections for r-process 
nucleosynthesis in the $^{48}$Ca region, namely for the isotopes
$^{40-44}$S, $^{46-50}$Ar, $^{56-66}$Ti, $^{62-68}$Cr, and 
$^{72-76}$Fe. While previously only cross 
sections resulting from the compound nucleus reaction mechanism 
(Hauser-Feshbach) have been considered, we recalculate not only that 
contribution to the cross section but also include direct capture on 
even-even nuclei. The level schemes, which 
are of utmost importance in the direct capture calculations, are taken 
from quasi-particle states obtained with a folded-Yukawa potential and 
Lipkin-Nogami pairing. Most recent deformation values derived from 
experimental data on $\beta$-decay half lives are used where available. 
Due to the consideration of direct capture, the 
capture rates are enhanced and the ``turning points'' in the r-process 
path are shifted to slightly higher mass numbers. 
We also discuss the sensitivity of the direct capture cross sections on 
the assumed deformation.
}

\vspace*{1cm}

\section{Introduction}

\renewcommand{\baselinestretch}{1.3}
\small \normalsize
A number of investigations (see e.g.~$^{1,2)}$) of the nuclear 
properties of neutron rich nuclei close to and at the magic neutron 
number $N=28$ has been motivated by the fact that these isotopes may 
play a crucial role in explaining the Ca-Ti-Cr-Fe isotopic anomalies 
found in meteoritic inclusions. Experimental information is 
scarce due to the short half-life of the involved nuclei. However, 
$\beta$-decay properties of several isotopes in that region 
have been measured~$^{1,3,4)}$ recently. Theoretical 
predictions can be improved by utilizing such experimental data.

\bigskip

For the isotopes $^{43}$P, $^{42}$S, $^{44}$S, $^{45}$S, $^{44}$Cl, 
$^{45}$Cl, $^{46}$Cl, and $^{47}$Ar, nuclear deformations have been derived 
from the measured $\beta$-decay 
half-lives in a recent QRPA parameter study~$^{3,4)}$. 
Taking these deformations we 
calculated the QP-levels below and close to the neutron separation 
energy, which are needed as input for the determination of the capture 
cross sections. However, we also extended our calculations to more 
neutron rich isotopes of S and Ar, and to Cr-, Ti- 
and Fe-isotopes by taking the deformations from a theoretical mass 
formula~$^{5)}$. With this level information we updated the theoretical 
cross sections of the even-even isotopes of $^{40-44}$S, $^{46-50}$Ar, 
$^{56-66}$Ti, $^{62-68}$Cr, and $^{72-76}$Fe. The importance of -- and 
the accuracy in the reproduction of experimental data by -- direct
capture for the 
case of $^{48}$Ca was already shown elsewhere~$^{6)}$.

\section{Method}
Mainly two reaction mechanisms have to be considered for astrophysically 
relevant 
neutron energies: the Compound Nucleus Mechanism 
(CN, Hauser-Feshbach, statistical model) and Direct Capture (DC) to 
bound states. For 
the majority of neutron-rich intermediate and heavy mass nuclei CN will 
dominate. However, the statistical model is only applicable as long as 
the level density is sufficiently high (i.e. $\ge 10$ MeV$^{-1}$). 
Therefore, DC may dominate for capture on nuclei with low level 
densities at the neutron separation energy. This can be the case for 
light nuclei, nuclei close to magic neutron numbers, and nuclei close to 
the neutron drip line (similar for proton capture on the proton rich 
side).

\bigskip

In previous r-process network calculations~$^{1)}$ only CN was considered in the 
theoretical neutron capture cross sections for the relevant nuclear 
region. Moreover, the level densities were computed from a back-shifted 
Fermi-gas formula~$^{7)}$. For our purposes we calculated 
quasi-particle levels in a folded-Yukawa potential and with 
Lipkin-Nogami pairing~$^{5)}$. The levels derived in such a way were 
used as input for the statistical model code SMOKER~$^{8)}$ and for the direct 
capture code TEDCA~$^{9)}$.

\bigskip

After the calculation of the nuclear energy levels and the CN cross 
sections, the DC contributions for capture on even-even nuclei (to first 
order one can assume that the level density in nuclei with one or two 
unpaired nucleons will be higher than for even-even nuclei, and that 
therefore DC will be most important for capture on even-even targets) 
were determined as follows. The theoretical cross section 
$\sigma^{\mathrm{th}}$ is given by a sum over each final state $i$~$^{10)}$
\begin{equation}
\label{dc}
\sigma^{\mathrm{th}}=\sum_i C_i^2 S_i \sigma_i^{\mathrm{ DC}} \quad.
\end{equation}
In our case the isospin Clebsch-Gordan coefficients $C_i$ are equal to 
unity. The 
spectroscopic
factors $S_i$ describe the overlap between the antisymmetrized wave 
functions of
target+n and the final state. In the case of one-nucleon capture on 
even-even deformed nuclei, the 
spectroscopic factor for capture into a state $i$, which has an occupation 
probability $v_i^2$ in the target, can be reduced to~$^{11)}$
\begin{equation}
S_i=1-v_i^2\quad.
\end{equation}
The corresponding probabilities $v_i^2$ are found by solving the 
Lipkin-Nogami pairing equations~$^{5)}$.

\bigskip

The factors $\sigma_i^{\mathrm{DC}}$ in Eq.~\ref{dc} are essentially 
determined by the overlap of the scattering wave function in the 
entrance channel, the bound-state wave function and the 
multipole-transition operator. The potentials needed for the calculation 
of the before-mentioned wave functions are obtained by applying the 
folding procedure. In this approach, the nuclear density of the target 
$\rho_{\mathrm T}$ is folded with an energy and density dependent 
effective nucleon-nucleon interaction $w_{\mathrm{eff}}$~$^{12)}$
\begin{equation}
V(E,R)=\lambda V_{\mathrm{F}}(E,R)=\lambda \int \rho_{\mathrm 
T}(\vec{r})w_{\mathrm{eff}}(E,\rho_{\mathrm T},\vert \vec{R}-\vec{r} 
\vert)\,d\vec{r}\quad,
\end{equation}
with $\vec{R}$ being the separation of the centers of mass of the two 
colliding nuclei. The interaction $w_{\mathrm{eff}}$ is only weakly 
energy dependent in the energy range of interest $^{13)}$. 
The density distributions $\rho_{\mathrm T}$ were 
calculated from the folded-Yukawa wave functions.

\bigskip

The only remaining parameter $\lambda$ was determined by employing a 
parametrization of the volume integral $I$
\begin{equation}
I(E)=\frac{4\pi}{A}\int V_{\mathrm{F}}(R,E)R^2\,dR\quad,
\end{equation}
expressed in units of MeV\,fm$^3$, and with the mass number $A$ of the 
target nucleus. Recently, the averaged volume integral $I_0$ 
was fitted to a function of mass number $A$, charge $Z$ and 
neutron number $N$ for a set of specially selected nuclei:~$^{14)}$
\begin{equation}
I_0=255.13+984.85A^{-1/3}+9.52\times 10^6 \frac{N-Z}{A^3}\quad.
\end{equation}
Thus, the strength factor $\lambda$ can easily be computed for each 
nucleus by using
\begin{equation}
\lambda=\frac{I_0}{I}\quad.
\end{equation}
For the bound states, the parameters $\lambda$ are fixed by the requirement 
of a correct 
reproduction of the separation energies.

\section{Results and Discussion}

The calculated CN and DC cross sections as well as the deformation 
parameters used for the calculation of the single-particle levels are shown in 
Table 1. The quoted neutron separation energies in the 
final nucleus are taken 
from an experimental 
compilation~$^{15)}$ where available, otherwise they were calculated from 
the theoretical mass formula~$^{5)}$. Furthermore,
$\beta$-decay half-lives are compared to theoretical lifetimes against 
neutron capture, computed from our results with a neutron number density 
of $3\times10^{19}$ cm$^{-3}$ (S, Ar) and $6\times10^{20}$ cm$^{-3}$ 
(Ti, Cr, Fe). Shown are experimental 
$\beta$-decay properties~$^{1,3,4)}$ and also theoretical values obtained 
by using
the QRPA code~$^{16)}$ with folded-Yukawa wave functions and
Lipkin-Nogami pairing for nuclei for which no experimental
$\beta$-decay properties were known.
\begin{table}
\renewcommand{\baselinestretch}{1.0}
\small \normalsize
\caption{Calculated 30 keV (c.m.) Maxwellian averaged neutron capture 
cross sections $<\sigma>_{30\,\mathrm{keV}}$ for CN and DC.
The column labeled `\%' gives the portion of direct capture in the total cross 
section. Also shown are the deformations $\epsilon_2$ and neutron 
separation energies 
S$_{\mathrm{n}}$ of the final nucleus target+n. The neutron capture 
half-lives T$_{1/2}$(n) were computed with the values from column 
`DC+CN' and a neutron number density of $3\times10^{19}$ (S, Ar) and 
$6\times10^{20}$ cm$^{-3}$ (Ti, Cr, Fe), respectively.}
\begin{center}
\begin{tabular}{clllllcll}
\hline
Target&\multicolumn{1}{c}{$\epsilon_2$}&\multicolumn{1}{c}{S$_{\mathrm{n}}$}
&\multicolumn{1}{c}{DC}&\multicolumn{1}{c}{CN}&\multicolumn{1}{c}{DC+CN}
&\multicolumn{1}{c}{\%}&\multicolumn{1}{c}{T$_{1/2}$(n)}
&\multicolumn{1}{c}{T$_{1/2}$($\beta$)}\\
 & &\multicolumn{1}{c}{[MeV]}&\multicolumn{1}{c}{[mb]}
&\multicolumn{1}{c}{[mb]}&\multicolumn{1}{c}{[mb]}
& &\multicolumn{1}{c}{[s]}&\multicolumn{1}{c}{[s]}\\
\hline
\hline
$^{40}$S&+0.24&3.8238$^{\ddagger}$&0.4246&0.0851&0.5097&83&0.218&8.60\\
$^{42}$S&+0.30$^*$&3.3114$^{\ddagger}$&0.9466&0.0202&0.9668&98
&0.115&0.56$\pm$0.06$^{\dagger}$\\
$^{44}$S&$-$0.20$^*$&1.3344&0.0000&0.0044&0.0044&0
&25.25&0.12$\pm$0.01$^{\dagger}$\\
$^{44}$S$^{\mathrm{a}}$&+0.30&1.3344&0.014&0.0044&0.0184&76
&6.040&0.12$\pm$0.01$^{\dagger}$\\
\hline
$^{46}$Ar&$-$0.18$^*$&4.2590$^{\ddagger}$&0.5295&0.1203&0.6498&81&0.171&7.80\\
$^{48}$Ar&$-$0.22&1.7074&0.0427&0.0144&0.0571&75&1.946&0.11\\
$^{50}$Ar&$-$0.28&1.0804&0.0016&0.0030&0.0046&35&24.15&0.05\\
\hline
$^{56}$Ti&+0.13&2.1936$^{\ddagger}$&0.0147&0.1305&0.1452&10&0.038&0.421\\
$^{58}$Ti&$-$0.10&2.4244&0.0185&0.0797&0.0982&19&0.057&0.152\\
$^{60}$Ti&$-$0.02&2.1194&0.0165&0.0403&0.0568&29&0.098&0.054\\
$^{62}$Ti&$-$0.04&0.5878&0.0068&0.0030&0.0098&69&0.569&0.018\\
$^{64}$Ti&+0.06&0.4094&0.0013&0.0002&0.0015&87&3.561&0.039\\
$^{66}$Ti&+0.14&0.3914&0.0009&0.0002&0.0011&82&4.960&0.013\\
\hline
$^{62}$Cr&+0.30&2.9134&0.0119&0.3846&0.3965&$<$1&0.014&0.349\\
$^{64}$Cr&+0.05&1.8614&0.0170&0.0353&0.0523&33&0.106&0.154\\
$^{66}$Cr&+0.10&2.2374&0.0126&0.0673&0.0799&16&0.071&0.071\\
$^{68}$Cr&+0.16&1.8274&0.0129&0.0268&0.0397&32&0.140&0.026\\
\hline
$^{72}$Fe&+0.14&2.2474&0.0101&0.0572&0.0673&15&0.083&0.089\\
$^{74}$Fe&+0.07&2.0564&0.0104&0.0360&0.0464&22&0.120&0.052\\
$^{76}$Fe&+0.06&0.0584&0.0050&$3\times10^{-6}$&0.0050&$>$99&1.12&0.045\\
\hline
\end{tabular}
\end{center}
$^{\mathrm{a}}$ $\epsilon_2$($^{45}$S)$=$$\epsilon_2$($^{44}$S)$=+0.3^*$ 
(see text)\\
$^*$ deformation inferred from experimental $\beta$-decay 
half-life~$^{3,4)}$\\
$^{\dagger}$ experimental $\beta$-decay half-life~$^{1,3,4)}$\\
$^{\ddagger}$ S$_{\mathrm{n}}$ from experimental values~$^{15)}$
\end{table}

\bigskip

From Table 1 one can see nicely the importance of direct capture when 
approaching the magic neutron number $N=28$ (S, Ar), but also the 
increasing contribution of DC to the cross section when approaching the 
drip line. The latter point can clearly be seen for the Ti isotopes and 
coincides well with the drop in level density at the neutron 
separation energy. The calculated Cr isotopes are still farther away 
from the drip line, therefore the capture cross section is dominated by 
CN.

\bigskip

An interesting effect can be seen for $^{44}$S: DC is suppressed totally 
when assuming a deformation $\epsilon_2$($^{45}$S)$=-0.2$, as suggested 
in order to reproduce the experimental $\beta$-decay 
half-life~$^{3,4)}$ of $^{45}$S. Due to the different deformations of 
target ($\epsilon_2$($^{44}$S)$=+0.3$) and final nucleus, the level 
order is changed and the previously unbound [303]7/2$^-$ level 
($E_x$=0.0582 MeV) is shifted well below the Fermi energy 
in $^{45}$S. This means that not only is a captured neutron added to one 
level of
$^{44}$S, but also that the nucleus is undergoing a reordering process. 
Such a process cannot be described by DC and therefore the respective 
cross section will vanish. As found in the QRPA parameter 
study~$^{3,4)}$, another deformation value consistent with the 
experiment would be $\epsilon_2$($^{45}$S)$=+0.125$. However, even with 
this value a similar effect can still be seen. Only with a deformation 
very close to the deformation of the target nucleus, non-zero DC cross 
sections can be obtained. Nevertheless, one has to note that the 
calculated QP levels might have an uncertainty which is larger than the 
distance of the level in question from the Fermi energy. Therefore, it 
might still be reasonable to calculate a DC contribution by assuming the 
same deformation for target and final nucleus.
The resulting cross sections for the two cases 
$\epsilon_2$($^{45}$S)$=-0.2$ and $\epsilon_2$($^{45}$S)$=+0.3$ are 
shown in Table 1.
The level schemes for $^{44}$S and $^{45}$S are shown in Table 2.
The astrophysical consequences are not changed for either deformation 
because the neutron capture lifetime is longer by orders of 
magnitude than the $\beta$-decay lifetime in both cases.
\begin{table}
\renewcommand{\baselinestretch}{1.0}
\small \normalsize
\caption{Quasi-particle levels for $^{44}$S and $^{45}$S.}
\begin{center}
\begin{tabular}{rl|rl}
\hline
\multicolumn{2}{c|}{$^{44}$S ($\epsilon_2=+0.3$)}&
\multicolumn{2}{c}{$^{45}$S ($\epsilon_2=-0.2$)}\\
\multicolumn{1}{c}{$E_{\mathrm{x}}$ [MeV]}&
\multicolumn{1}{c|}{[Nn$_{\mathrm{z}}$$\Lambda$]$\Omega$}&
\multicolumn{1}{c}{$E_{\mathrm{x}}$ [MeV]}&
\multicolumn{1}{c}{[Nn$_{\mathrm{z}}$$\Lambda$]$\Omega$}\\
\hline
\hline
$-$27.1979&    [ 0  0  0] 1/2&$-$26.0041&    [ 0  0  0] 1/2\\
$-$20.3364&    [ 1  1  0] 1/2&$-$18.2826&    [ 1  0  1] 3/2\\
$-$17.8157&    [ 1  0  1] 3/2&$-$17.6452&    [ 1  1  0] 1/2\\
$-$15.8641&    [ 1  0  1] 1/2&$-$15.5487&    [ 1  0  1] 1/2\\
$-$12.4255&    [ 2  2  0] 1/2&$-$10.3816&    [ 2  0  2] 5/2\\
$-$10.3579&    [ 2  1  1] 3/2&$-$9.2965&    [ 2  1  1] 3/2\\
$-$7.8438&    [ 2  0  2] 5/2&$-$8.9932&    [ 2  2  0] 1/2\\
$-$7.5158&    [ 2  0  0] 1/2&$-$5.8346&    [ 2  0  0] 1/2\\
$-$4.1707&    [ 2  1  1] 1/2&$-$5.7213&    [ 2  0  2] 3/2\\
$-$3.9511&    [ 3  3  0] 1/2&$-$3.0690&    [ 2  1  1] 1/2\\
$-$3.2493&    [ 2  0  2] 3/2&$-$1.7317&    [ 3  0  3] 7/2\\
$-$2.3583&    [ 3  2  1] 3/2&$-$0.6623&    [ 3  1  2] 5/2\\
$-$0.5604&    [ 3  1  2] 5/2&$-$0.3329&    [ 3  2  1] 3/2\\
0.0000&    [ 3  1  0] 1/2&$-$0.1400&    [ 3  3  0] 1/2\\
0.0582&    [ 3  0  3] 7/2&0.0000&    [ 3  0  1] 3/2\\
0.9147&    [ 3  2  1] 1/2&0.5831&    [ 3  1  0] 1/2\\
&&1.2550&    [ 3  0  3] 5/2\\
\hline
\end{tabular}
\end{center}
\end{table}

\section{Summary}

It has been shown that it is important to consider DC cross 
sections as well as CN cross sections for nuclei close to magic numbers 
and close to the drip lines. However, in order to reliably calculate DC 
contributions one needs level schemes which can be 
calculated from microscopic models. In this context, improved experimental 
data is of utmost importance not only for a better knowledge of 
$\beta$-decay half-lives, but also to be able to infer deformation parameters.
With our new neutron capture cross sections we obtain ``turning points'' 
in the r-process path for the mentioned neutron densities at $^{44}$S, 
$^{48}$Ar, $^{62}$Ti (with a minor branching at $^{60}$Ti), $^{68}$Cr 
($^{66}$Cr), and $^{74}$Fe ($^{72}$Fe). (Note, however, that the 
theoretical $\beta$-decay half-lives shown are taken from QRPA 
calculations~$^{16)}$. Other models might yield different values).
Full reaction network 
calculations with varying neutron densities have to be performed 
for a more detailed 
determination of the resulting abundances and isotopic ratios.

\medskip

{\bf Acknowledgement:} We are indebted to 
P. M\"oller for discussions 
and for making his QRPA code available to us. We also thank F.-K. 
Thielemann for discussions. This work was supported in part by the 
Austrian Science Foundation (S7307--AST), the \"Osterreichische
Nationalbank (project 5054) and the DFG (Kr 806/3).

\section*{References}

\renewcommand{\baselinestretch}{1.0}
\small \normalsize
\newcounter{tomref}
\begin{list}{$^{\arabic{tomref})}$}{\usecounter{tomref}
\itemsep0cm
\parsep0cm}
\item O. Sorlin et al., {\it Phys.\ Rev.\ C} {\bf 47} (1993) 
2941;\\
O. Sorlin et al., {\it Proc. Nuclei in the Cosmos III}, eds.\ M. Busso, 
R. Gallino, C.M. Raiteri (AIP Press, New York 1995), p.\ 191.
\item K.-L. Kratz, A.C. M\"uller, F.-K. Thielemann, {\it Phys.\ 
Bl.} {\bf 51} (1995) 183.
\item W. B\"ohmer, P. M\"oller, B. Pfeiffer, K.-L. Kratz, in {\it 
IKCMz report}, ed.\ H. Denschlag (IKCMz, Mainz 1995), p.\ 35.
\item W. B\"ohmer et al., to be published.
\item P. M\"oller, J.R. Nix, W.D. Myers, W.J. Swiatecky, {\it 
At.\ Data Nucl.\ Data Tables} (1995), in press.
\item A. W\"ohr et al., in {\it Proc. VIII. Int. Symp. on Gamma-Ray 
Spectroscopy and Related
     Topics, Fribourg, Schweiz}, ed.\ J. Kern (World Scientific:
     Singapore 1994), p.\ 762.
\item J.J. Cowan, F.-K. Thielemann, J.W. Truran, {\it Phys.\ 
Rep.} {\bf 208} (1991) 267.
\item F.-K. Thielemann, M. Arnould, J.W. Truran, in {\it 
Advances in Nuclear Astrophysics}, eds.\ E. Vanghioni-Flam et al.,
(\'editions fronti\`eres, Gif-sur-Yvette 1987), p.\ 525.
\item H. Krauss, unpublished.
\item K.H. Kim, M.H. Park, B.T. Kim, {\it Phys.\ Rev.\ C} {\bf 23} 
(1987) 363.
\item N.K. Glendenning, {\it Direct Nuclear Reactions} (Academic Press, 
New York 1983).
\item A.M. Kobos, B.A. Brown, R. Lindsay, G.R. Satchler, {\it Nucl.\ 
Phys.} {\bf A425} (1984) 205.
\item H. Oberhummer, G. Staudt, in {\it Nuclei in the Cosmos}, ed.\ H. 
Oberhummer (Springer, Berlin 1991), p.\ 29.
\item W. Balogh, diploma thesis, University of Technology, Vienna 
1994;\\
W. Balogh et al., to be published.
\item G. Audi, A.H. Wapstra, {\it Nucl.\ Phys.} {\bf A565} (1993) 1.
\item P. M\"oller, J. Randrup, {\it Nucl.\ Phys.} {\bf A514} (1990) 1.
\end{list}
\end{document}